\documentclass[runningheads]{llncs}
\usepackage[T1]{fontenc}
\usepackage{graphicx}
\usepackage{booktabs}
\usepackage{array}
\usepackage{paralist}

\newcolumntype{C}[1]{>{\centering\arraybackslash}p{#1}}
\newcolumntype{L}[1]{>{\arraybackslash}p{#1}}
\usepackage{hyperref}
\usepackage{xcolor}
\usepackage{color}

\usepackage{academicons}
\definecolor{orcidlogocol}{HTML}{A6CE39}

\usepackage{microtype}

\begin{document}
\title{Aspect-Driven Structuring of \\ Historical Dutch Newspaper Archives}

\author{Hermann Kroll\inst{1}\orcidID{0000-0001-9887-9276} \and
Christin Katharina Kreutz\inst{2}\orcidID{0000-0002-5075-7699} \and
Mirjam Cuper\inst{3}\orcidID{0000-0003-0187-9873} \and
Bill Matthias Thang\inst{1} \and
Wolf-Tilo Balke\inst{1}\orcidID{0000-0002-5443-1215}
}
\authorrunning{Kroll et al.}
\institute{TU Braunschweig, Germany, \email{\{kroll, balke\}@ifis.cs.tu-bs.de}\and
TH Köln (University of Applied Sciences), \email{christin.kreutz@th-koeln.de}\and
KB, National Library of the Netherlands, The Netherlands, \email{mirjam.cuper@kb.nl}}
\maketitle              % typeset the header of the contribution
\begin{abstract}
Digital libraries oftentimes provide access to historical newspaper archives via keyword-based search. Historical figures and their roles are particularly interesting cognitive access points in historical research. Structuring and clustering news articles would allow more sophisticated access for users to explore such information. However, real-world limitations such as the lack of training data, licensing restrictions and non-English text with OCR errors make the composition of such a system difficult and cost-intensive in practice. In this work we tackle these issues with the showcase of the National Library of the Netherlands by introducing a role-based interface that structures news articles on historical persons. In-depth, component-wise evaluations and interviews with domain experts highlighted our prototype's effectiveness and appropriateness for a real-world digital library collection.

\end{abstract}

\section{Introduction}
Users of digital libraries featuring historical news articles conduct a variety of information interactions such as task planning or searching for and working with information objects~\cite{DBLP:journals/jd/LateK22}.
In historical research, historical figures and especially their roles are particularly interesting cognitive access points~\cite{DBLP:journals/jasis/KumpulainenKZS20}.
Kumpulainen et al.~\cite{DBLP:journals/jasis/KumpulainenKZS20} state the need for supporting historians' research by providing domain-specific tools tailored to their needs.
One crucial task of researchers is the creation of sub-corpora to answer their research questions~\cite{DBLP:journals/jdmdh/PfanzelterOMLH21}.
However, finding these sub-corpora, especially when researchers are unfamiliar with the searched historical persons, can be challenging for two reasons.
First, the huge size of news article archives might be overwhelming. 
Second, posing and finding suitable keyword queries to browse such archives is difficult. 

Advances in Natural Language Processing (NLP) lead to historic news systems with novel access paths for users to engage with their content~\cite{histNewspaperReview}. 
A variety of such digital library projects has been proposed in the past, e.g., NewsEye~\cite{NewsEye}, ANNO~\cite{ANNO},  impresso~\cite{impresso}, or Cuper's work~\cite{DBLP:conf/ercimdl/Cuper21} (see Sect.~\ref{sec:related_work} for a detailed discussion). However, those systems usually rely either on manual curation~\cite{DBLP:conf/ercimdl/Cuper21} or at least domain-specific training examples for every implemented step~\cite{impresso}.
In contrast, our work bypasses manual curation and the collection of domain-specific training data by utilizing data from Wikipedia (structure information with text examples).  
This paper introduces a novel system that automatically structures historical news articles on persons and provides an aspect-driven interface to explore a library's content. 
The central idea is that a person has different roles (e.g., \textit{writer}, \textit{politician}, \textit{military person}) and each role has different aspects (e.g., \textit{early life, political career, actions}). 
Our system should, at best, automatically create sub-corpora for each role and aspect to support research on historical persons. 
However, traditional methods introduced in the NLP domain typically rely on hand-crafted training data and sometimes artificial benchmarks~\cite{PRIMERA}. 
We tackled the challenges faced by an actual digital library, namely the National Library of the Netherlands, Koninklijke Bibliotheek (KB)~(\url{https://www.kb.nl}). 
Here, no hand-crafted training data and benchmarks were available.
Moreover, the library imposed several real-world constraints: 
(1) The data was available in Dutch, whereas NLP methods are often available in English only.
(2) The news articles were based on OCR-scanned newspapers, and hence, came with typical OCR issues (such as incorrect letters or broken paragraphs). 
(3) The data came with a license prohibiting sending data to APIs like ChatGPT~\cite{chatgpt}. 
%Hence, data had to be processed non-publicly, and methods had to be scalable for an actual application. 

In addition to those constraints, which are typically not the target in NLP research, we observed an understudied~\cite{DBLP:journals/jd/LateK22} corpus of non-English but Dutch news articles. 
Our overall goal was thus to build a real-world system that overcomes the typical constraints of a typical digital library. 
In this work, we therefore strive to support users' data-driven process planning by structuring news articles concerning historical figures by their respective roles.  
%(see Fig.~\ref{fig:system_overview}).\todo{c thinks: include more detail in figure, "person class" is not mentioned in the paper. Cutting snippets from newspapers, translating and summarizing is not contained. A user formulating a query is not part of our interface (also misleading as we only provide a number of persons). The figure contains "narrative structuring" what is the narrative?}
Our prototypical system operates on real data of the KB and bases on automatically generated training data from Wikipedia. 
We expected our system to help users in the formulation of research questions on the provided data of historical persons. 

To tackle our overall research question \textit{How can a digital library design effective access paths to explore their collection?}, we made the following contributions: 
(1) We discuss and demonstrate how we overcome a digital library's real-world restrictions and constraints (see Sect.~\ref{sec:restrictions}). 
(2) We present an effective method for automatically structuring news articles by employing structural background information from Wikipedia with the use case of news articles on historical figures. 
(3) We evaluate our prototypical system step-by-step and via interviews with five domain experts.
Code is available at GitHub\footnote{\url{https://github.com/HermannKroll/AspectDrivenNewsStructuring}} and Software Heritage\footnote{\url{https://archive.softwareheritage.org/swh:1:dir:13457c154ed7ad1f571e353c1edf2f87db61b0ae}}. 

%summarize and discuss the capabilities and limitations

\section{Related Work}
\label{sec:related_work}
Related work for our research objective falls into the following categories: (1) related digital library news archive retrieval systems, (2) processing Dutch texts via language models, and (3) text summarization methods.

\textit{Digital Library Systems on News Articles.}
Structuring and exploring news has been a topic of wide research, e.g., summarization~\cite{PRIMERA}, the evolution of terms~\cite{DBLP:conf/histoinfo/MarjanenPZK19}, fake news detection~\cite{DBLP:conf/ercimdl/VogelJ19}, clustering~\cite{DBLP:conf/ercimdl/MariaS00} and many more.
For instance, \cite{DBLP:conf/ercimdl/MariaS00} clusters news articles based on their similarity to pre-computed categories using SVMs.
Kumpulainen et al.~\cite{DBLP:journals/jasis/KumpulainenKZS20} identified roles of historical persons, relationships between them, and in general, named entities as important cognitive access points to historical documents.
Clustering similar news articles has been explored in several concrete applications with real digital library constraints, e.g., NewsEye~\cite{NewsEye} or ANNO~\cite{ANNO}.
Another example is the Swiss-Luxembourgish project impresso~\cite{impresso} which utilizes NLP methods like named entity recognition, word embeddings, n-gram search, and information extraction to provide additional information on historical news articles. 
The KB has developed the Delpher platform: 
News articles were digitized by OCR tools and Delpher provides a user interface to navigate through their historical newspaper collections.
Beyond the traditional keyword-based search, they aimed to organize a part of the KB's newspaper collection differently from the standard search interface~\cite{DBLP:conf/ercimdl/Cuper21}.
Additionally, the KB manually created subject pages that give more background information on certain topics and related newspapers\footnote{\url{https://www.delpher.nl/thema/geschiedenis/tweede-wereldoorlog}}. 
%However, the creation of such pages required much manual work.
Our work's goal was to structure the KB's news articles automatically, at least as much as possible, while meeting the KB's real-world constraints.

\textit{Dutch Language Models.}
Many language models were trained and evaluated on English corpora.
Exceptions were models trained in a \textit{multilingual setting}~\cite{BERT,mBART,mt5} or ones having been \textit{trained for Dutch}: 
BERTje~\cite{bertje} is a Dutch BERT~\cite{BERT} model which outperforms the multilingual version~\cite{mBART} of BART~\cite{BART}. 
RobBERT~\cite{robbert} is a Dutch RoBERTa model which outperformed BERTje on the sentiment analysis task as well as both BERTje and mBERT on the relative pronoun prediction tasks. 
A newer version of the model (RobBERT-2022)~\cite{RobBERT22} with a newer Dutch training corpus also outperformed BERTje and RobBERT on the sentiment analysis task.
We used the RobBERT-2022 for our text classification.

%On named entity recognition and part-of-speech tagging they found mBERT~\cite{BERT} surpassing both BERTje and RobBERT.
% on word-level NLP tasks, such as named entity recognition, part-of-speech tagging or semantic role labelling.
%

%
%\subsubsection{Multilingual Language Models.}
%mBERT~\cite{BERT} is the multilingual version of BERT with 110M parameters. The model was trained on 104 languages, including Dutch. The model was shown to perform well on Dutch NLP tasks, such as named entity recognition~\cite{DBLP:conf/emnlp/WuD19}.
%
%mBART~\cite{mBART} is the multilingual version of BART~\cite{BART}. It has been trained on 25 languages, including Dutch.
%
%mT5~\cite{mt5} is the multilingual version of T5. 1.98 percent of its training data (96 million pages) was in Dutch. In an entity recognition task the large model (and bigger models) outperforms mBERT.
%
%
%\subsubsection{Dutch Language Models.}

\textit{Text Summarization.} 
The task of text summarization is to produce a concise natural language summary. 
Nowadays, general-application sequence-to-sequence language models can be fine-tuned to solve the text summarization task, e.g., UniLM~\cite{UniLM}, T5~\cite{t5}, BART~\cite{BART}, PEGASUS~\cite{PEGASUS}. 
Another option is using large language models (LLMs)~\cite{DBLP:journals/corr/abs-2301-13848} and prompting in this context.
Models like \cite{PEGASUS} or \cite{BART} are restricted to 512 tokens or less, meaning their input must be shorter than 512 tokens. 
So-called longformer models surpass this restriction by allowing up to 16k tokens as their input, e.g.,
\cite{BigBird,LED,PEGASUS-X}. 
Beyond some remarkable examples like Estonian~\cite{DBLP:journals/bjmc/HarmA22} and Romanian news summarization~\cite{DBLP:journals/algorithms/NiculescuRD22}, text summarization models are trained in English~\cite{t5,BART,PEGASUS,LED,PRIMERA}, but they can be fine-tuned for other languages (here Dutch). 
The goal of this work was to summarize several articles in a single summary, so the \textit{multi-document summarization} task. %, i.e., generate a single summary for several documents. 
PRIMERA~\cite{PRIMERA} is an LED-based~\cite{LED} state-of-the-art model (ACL2022) for this task. 
It outperformed single-document models~\cite{PEGASUS,BART,LED} in different scenarios (news and scientific documents).
That is why we used PRIMERA in this work.

\section{Conception and Data Acquisition}
Our overall goal was to structure news articles to support corresponding research questions on single persons. 
Each new article consists of a title, the textual content, the release date, and the publishing newspaper. 
From our viewpoint, each person might have different roles $r \in \mathcal{R}$ (e.g., politician, writer, artist) that come with different aspects $r_A$ (e.g., political career, novels, awards).

\begin{figure*}[t]
\centering
\includegraphics[trim=2.0cm 1cm 0.0cm 2.0cm, width=0.9\textwidth]{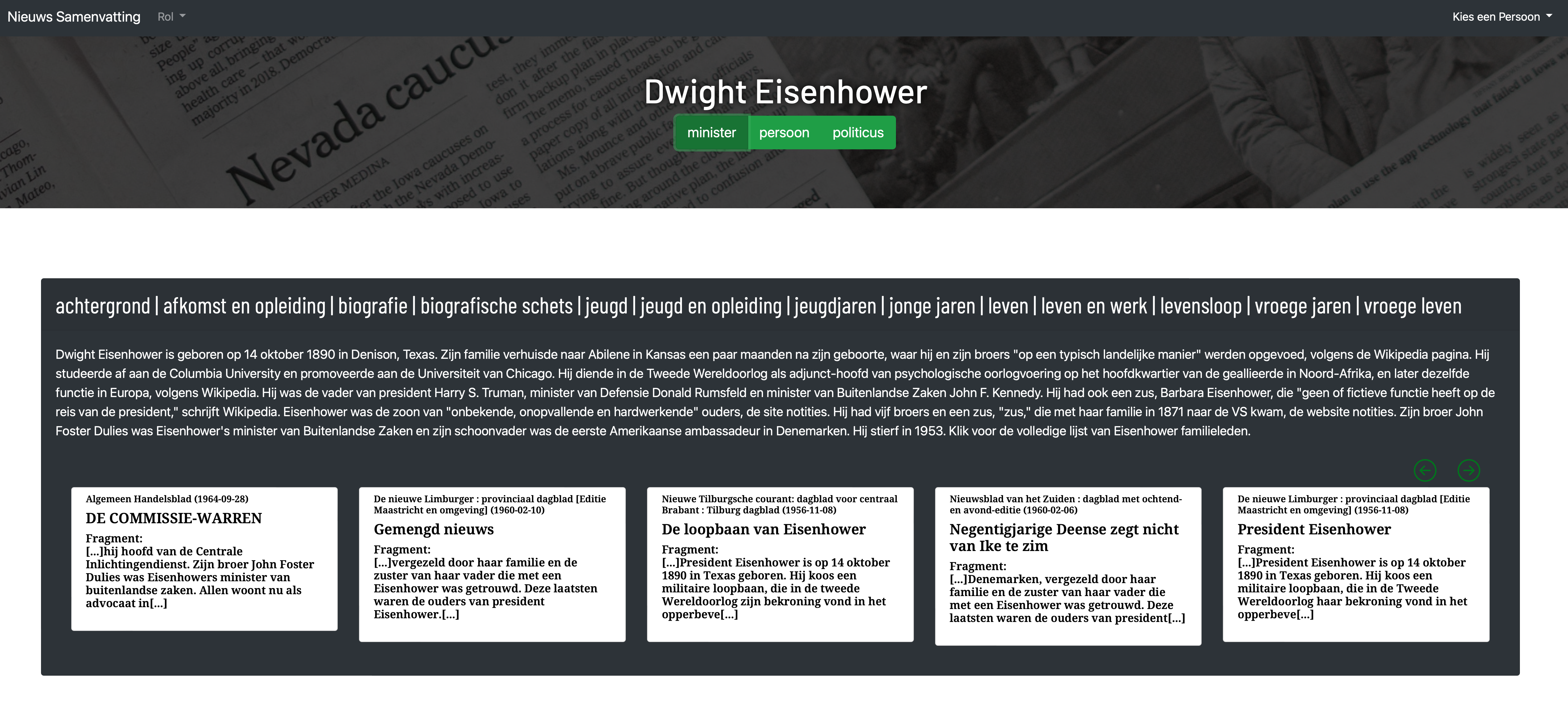}
\caption{User interface of our system, URL: \url{https://narrative.pubpharm.de/news}.}
\label{fig:screenshot}
\end{figure*}

\textit{Discussion of the Library's Constraints.}
\label{sec:restrictions}
In brief, we faced the following constraints: 
(1) The texts stem from OCR-scanned news articles using ABBYY Finereader,  
(2) texts were written in Dutch (no translation was available),
(3) prohibition against sending data to third parties,
(4) forced linking to the Delpher system and restriction to show only snippets of the actual data (160 characters at max), and
(5) no curated training data for any of our sub-tasks. 
Those requirements forced us to exclude automated translation services like DeepL and AI assistants like ChatGPT by design.
Especially the lack of training data prevented the usage of straightforward approaches like training text classification models. 
We would have had to collect data for roles and aspects, manually label news articles, and then train classification models. 
However, creating such data would be cost intensive. 

That is why we headed for a different approach:
We used the Dutch Wikipedia to gather texts describing different persons, their roles, and the roles' aspects. 
First, Wikipedia organizes text into different sections describing different \textit{aspects} of entries. 
Second, Wikipedia enriches an item's text through so-called info boxes that provide structured information, e.g., whether it is a person and has some roles. 
For our approach, we used the info boxes to determine a person's role and the Wikipedia texts to learn how certain aspects are described.  
This approach bypassed the creation of training data, while, however, could cause new problems: 
It had to be tested if classifiers trained on descriptive Wikipedia texts are transferable and generalizable to Dutch news.

\textit{Prototype (User Interface).} 
In constructing the system interface, we strive to carter to McCay-Peet et al.'s~\cite{DBLP:journals/ires/McCay-PeetTK14} five facets supporting serendipity in digital environments: interfaces filled with various information (\textit{trigger-rich}), showcase relationships between information objects (\textit{enables connections}), visual cues (\textit{highlights triggers}), \textit{enables exploration} and provides unanticipated or surprising information (\textit{leads to the unexpected}). 
Our method's goal was to (1) derive the roles of a person (trigger-rich and exploration) and (2) classify whether a news article's content belongs to one of the role's aspects (connections and unexpected). 
We used multi-document summarization for each aspect to help users quickly access what is written in the corresponding articles.
Users should be able to select different persons and one of the person's known roles. 
Then users could navigate through different aspects of that role, see a summary for each aspect and a list of articles classified as belonging to that role's aspect (see Fig.~\ref{fig:screenshot} for the systems' screenshot and URL).
A click on an article forwards users to Delpher.

\textit{Historical News Data from the KB.}
We used a subset of the KB's data for building our system since the KB collected news articles from the 17th century to the recent past.
We selected articles for nine famous persons in relation to the second world war with various roles because the KB's Delpher has identified the second world war as a topic users were interested in. 
We harvested relevant articles by querying for the name and title/pseudonym (see Tab.~\ref{tab_persons} for statistics). 
We only selected items from the newspaper collection with the type 'article'.
Then we only kept articles where $\ge$ 90\% of the text was found in a Dutch dictionary, as recommended in~\cite{ocrquality}, to remove noisy and low-quality OCR-scanned data. 
We also excluded newspapers published by fascist organizations or German authorities with a national socialist agenda.  
Articles for each person should, on the one hand, carry enough information about the person and, on the other hand, stem from the time when the person was alive.
That is why we applied the following additional filters:
(1) A news article's release date must be in the corresponding person's life span,  
(2) the article content must be longer than 100 words, and 
(3) a person's partial name (e.g., \textit{Frank} for \textit{Anne Frank}) should be mentioned at least three times. 
Especially the time constraint did filter nearly all articles, except one, of Anne Frank because they were published after her death.

\begin{table}[t]
\centering
\caption{Statistics for our news article collection: \#Art. describes how many articles before filtering and \#FArt. after filtering were retrieved for each person.}
\begin{tabular}{L{0.5\textwidth}rcrr}
\textbf{Person Name} (\textbf{Title/Synonyms})  & \textbf{Role} & \textbf{Life} & \textbf{\#Art.} & \textbf{\#FArt.} \\
\toprule
Winston Churchill (Sir Churchill) & politician  & 1874-1965 & 47k &  8463 \\

Leopold III van Beglië  (Leo. III, prins Leo.)  & king & 1901-1983 & 26k  & 1677  \\       
Wilhelmina van Oranje-Nassau (prinses Wilhelmina, koningin Wilhelmina)  & queen  & 1880-1962 & 257k   & 9416 \\  

Jannetje Schaft (Hannie Schaft)  & resistance  & 1920-1945 & 2056  & 34 \\ 
Dwight Eisenhower (majoor-generaal E., Generaal E., president E.) & politician   & 1890-1969 & 114k &  21k  \\ 

Anne Frank     & war victim   &  1929-1945 &  11k &  1 \\  

Frans Goedhart (Pieter 't Hoen)   & resistance   & 1904-1990 & 4105  &  560  \\  

Simon Vestdijk  & writer    & 1898-1971 & 5544  & 1453    \\   

Franklin Roosevelt (president Roosevelt) & politician & 1882-1945 & 165k & 16k  \\
\end{tabular}
\label{tab_persons}
\end{table}

\section{System Implementation}
\textit{Wikipedia Processing.}  
As already mentioned, we used the Wikipedia info boxes to derive a person's role. 
The information was linked to Wikipedia categories which were organized in a taxonomy, e.g., \textit{British politician} is a specialization of a \textit{politician}. 
In our context, we understood a person's occupation as a role.
We crawled the Dutch \textit{occupation} categories and derived a list of occupations (in sum 30k distinct ones).
Then, we iterated through the Dutch Wikipedia XML dumps (March 2023), parsed the info boxes, checked whether a property of the info box was linked to one of those occupations, and if so, we extracted the corresponding page's summary (introduction) and sections plus all occupations. 
In sum, we derived 259k person pages. 
While reviewing the pages, we observed many very short pages, e.g., including a brief summary or a single section.
However, our goal was to find frequent aspects of well-described roles. 
So, we removed all pages that (1) had a less than 150 characters summary, or (2) had $<$ 3 sections.
Note that we disregarded sections with less than 100 characters and sections that only contained references/literature by using a hand-crafted list.  
This filtering reduced the number of person pages to 61k. 
With that, we obtained roles plus thousands of Wikipedia pages for each.
Wikipedia sections should, at best, describe one unit of information belonging to a certain aspect of a person. 

However, Wikipedia was crafted collaboratively, i.e., through human editing meaning section titles are usually not-canonicalized. 
For instance, \textit{life}, \textit{background}, and \textit{curriculum vitae/resume} describe the same, or at least a very similar, aspect of a person.
To face this concern, we designed a canonicalization step to cluster semantically similar sections. 
We applied a pre-trained sentence transformer model (BERT-base-dutch-cased) using the S-BERT Library~\cite{reimers-2019-sentence-bert}, capable of embedding semantically similar sentences closely in its vector space. 
To embed a section, we embedded all of its sentences and then computed the mean vector over all sentence vectors.
Next, we averaged all section vectors with equal titles (e.g., background sections). 
Finally, we compared those sections vectors pairwise using the cosine similarity.
If two vectors' similarity exceeds a certain pre-defined threshold, we consider those section titles as semantically equivalent. 
We then computed the transitive closure to determine the set of semantically equivalent section titles, i.e., if a-b and b-c are merged, we also merged a-c. 
To retain a high precision, we used a  similarity threshold of 0.95 in our system. 

\textit{Aspect Mining and Classification.} 
Next, we mined frequent role aspects by counting how often the aspect (section title or any other section title from that same cluster) was used across all persons of a role (e.g., \textit{writer}s). 
We then computed a relative support, e.g., 0.2 means that 20\% of all \textit{writers} have aspect (section tile or any similar title) \textit{background}.
We defined a minimum absolute (to ensure enough text examples for aspect training) and a relative support threshold (to ensure frequency within a role). 
Given a certain person's role, we trained a classifier to predict whether a text belongs to one of the role's aspects.
That means we headed for a multi-class classification scenario, e.g., a classifier for role \textit{$r_1$} with aspects \textit{$a_1$, $a_2$, $a_3$} must predict one of the aspects, or the negative class (not belonging to the role). 
First, we retrieved Wikipedia section texts for each aspect.
We ensured that each aspect must have at least a minimum number of texts to be considered for training (see aspect mining support threshold).
However, some aspects might have more examples than others, which is why we sampled all text examples randomly down to the number of the least frequent aspect, e.g., aspect \textit{$a_1$} and  \textit{$a_2$} are sampled down to 100 texts if the least frequent aspect \textit{$a_3$} only has 100 examples.
We randomly sampled negative examples (not belonging to the role) from other persons and aspects that do not have the given role $r_1$.
We sampled as many negative examples as we had positive ones, e.g., 100-100-100 positive (three aspects, 100 texts each) and 300 negative examples. 

We fine-tuned the Dutch model RobBERT-2022~\cite{RobBERT22} for the actual text classification.
We split our data into train, validation, and test sets (80-10-10).
We performed training on train (5 epochs), and searched for hyperparameters (learning rate [1e-3, 1e-4, 1e-5] and decay [0.1, 0.2]) on validation.
We picked the best model concerning validation and macro precision  because our classification should prefer precision of all classes over recall. 
We trained a classifier for each role (occupation category of Wikipedia) that had (1) at least three frequent aspects and (2) belongs to the first two category levels in Wikipedia (to select more general roles like \textit{writer} instead of \textit{British writer}). 
Note that we removed the category suffixes \textit{naar nationaliteit} and \textit{naar beroep}.

\textit{News Article Processing.}
The next step was applying those classifiers to a historical person's actual Dutch news articles. 
However, a news article might include several different topics, thus, classifying the whole text to one role's aspects could be problematic. 
So, we computed snippets of the articles that include the person's name: 
We split the article's content into sentences by using NLTK's~\cite{DBLP:conf/acl/Bird06} sentence split method. 
We then checked whether a partial name of the person (e.g., \textit{Churchill} or \textit{Winston} for \textit{Winston Churchill}) was included. 
If so, we considered the sentence relevant and took it and the sentence before and after as additional context information to generate a snippet.
The average sentence length computed over Dutch newspaper and Wikipedia articles is 90.3 characters, 3.4\% of these sentences have $\le$ 19 characters~\cite{Dutch_word_length}. 
We only use snippets of three sentences with at least 50 characters to filter out broken or incomplete sentences, corresponding to a minimum average sentence length of $\sim$16.7 characters each.

For our selected persons, we identified their roles through the info boxes of the corresponding Wikipedia entry. 
If a role (e.g., \textit{British minister}) was assigned, we also considered its super categories (e.g., \textit{minister} and \textit{person}). 
We always assigned the role \textit{person} to ensure that our approach also worked in cases when a person did not have an info box (in cases of \textit{Wilhelmina} and \textit{Janeetje Schaft}). 
Having the roles, we applied the corresponding role classifier to every news article snippet of the person. 
Note that each snippet could be classified as belonging to several aspects of different roles. 
This was intended because some aspects of different roles might overlap, e.g., a \textit{politician}'s and \textit{writer}'s \textit{family} or \textit{early life}.

Our last goal was to summarize those snippets into one summary for the users so that they could quickly grasp how the aspect was described in the news articles. 
However, to the best of our knowledge, multi-document summarization models were unavailable for Dutch.  
That is why we decided to apply one of the latest English models, namely PRIMERA~\cite{PRIMERA}. 
We used a fine-tuned news summary PRIMERA model from HuggingFace. 
However, to apply PRIMERA, we had to translate the Dutch news article snippets into English with OPUS-MT~\cite{opusmt}, one of the latest open available translation models.
The choice of OPUS-MT over using, e.g., the DeepL API, was again made due to our legal constraints. 
Afterward, PRIMERA's English summaries were translated back to Dutch with OPUS-MT. 
We translated Dutch texts sentence-wise to English and vice versa. 
To generate the summaries, we introduced a parameter \textit{k} to select how many articles snippets should be summarized.
In addition to the summaries, we wanted to display fragments of the article snippets in the user interface, to give our users an idea about the article.
For the fragment generation, we identified the position where the person's name was mentioned  and displayed the surrounding characters and cut if we exceed the 160 characters we were allowed to show.

\section{Evaluation}
We evaluate our system's components individually (clustering, classification, translation, and summarization) and then report our user study's findings.

\textit{Clustering.}
We exported 221 distinct section titles that occurred in at least 100 Wikipedia articles to ensure enough examples for the clustering and classification.
We asked three persons to cluster them manually, i.e., whether two titles semantically belong together. 
When comparing and discussing their clusters, we observed the following patterns: 
% different level of detail possible
% clustering by different life periods (youth, late years/death) or everything on life into one
There was a wide range in clustering regarding the granularity.
One annotator clustered everything belonging to one's life as one cluster, whereas a second person created clusters for different periods in life such as \textit{youth} with \textit{early life}, \textit{youth and training} and \textit{later life} with \textit{death} and \textit{last years}.
 % difficult to differentiate between a person and the work of a person:
% - cluster about running/sports
%- music cluster with either discography, instruments usage
%- author, novel
%- actor, film
%- counter example: politician, political career
The annotators had difficulties distinguishing between titles describing a person, e.g., \textit{author}, and titles describing a person's work, e.g., \textit{novel}. 
But all annotators differentiated between a \textit{politician} and their \textit{political career}.
% work as a cluster
All three annotators agreed to cluster section titles describing %some form of \textit{work}, e.g., \textit{oeuvre} and 
different types of \textit{awards}. 
% differentiating on different careers: military, political
The annotators disagreed on whether to cluster military and political careers.
% wars together, interbellum, after war difficult
War-related titles such as \textit{interbellum} and \textit{after the war} also were regarded with uncertainty regarding them being in separate or the same cluster.
%  some things cannot be clustered, based on the title, it is sometimes problematic: work could be about the labor/job or the outcome of the role
In general, the annotators found that some section titles were very hard to cluster as the titles were ambiguous: 
\textit{Work} could be associated with a person's job, but also with its outcome, e.g., paintings of a painter.

\begin{table}[t]
\centering
\caption{Evaluation results for our Wikipedia text classifiers. We averaged the number of trained aspects and used training samples. Evaluation metrics are macro averaged.}
\begin{tabular}{lcccccc}
\label{tab_eval_classification}
\textbf{Setting} & \textbf{\#Aspects} & \textbf{\#Samples} &  \textbf{*Precision} & \textbf{Recall} &  \textbf{F1} & \textbf{Accuracy}  \\
\toprule
Top-5 & $7.6\pm3.83$ &  $9999\pm9520$ & $0.95\pm0.01$ & $0.94\pm0.03$ & $0.94\pm0.02$ & $0.95\pm0.02$ \\
Top-10 & $6.7\pm3.16$ &  $10246\pm14078$ & $0.94\pm0.02$ & $0.93\pm0.02$ & $0.93\pm0.02$ & $0.94\pm0.02$ \\
Worst-5 & $6.4\pm1.74$ &  $1285\pm230$ & $0.80\pm0.02$ & $0.79\pm0.03$ & $0.78\pm0.03$ & $0.82\pm0.02$ \\
Worst-10 & $5.7\pm1.95$ &  $1545\pm1340$ & $0.81\pm0.02$ & $0.81\pm0.04$ & $0.80\pm0.03$ & $0.83\pm0.02$ \\
\midrule
All (43) & $6.35\pm2.88$ &  $4254\pm8215$ & $0.87\pm0.05$ & $0.88\pm0.05$ & $0.87\pm0.05$ & $0.89\pm0.04$ \\

\end{tabular}
\end{table}

In a subsequent discussion, the three annotators also reviewed the system-generated clusters (41 in total) and commented on them. 
The annotators were content with most of the clustering but found some clusters which they considered too broad 
% computer generated clusters sometimes too broad: werken, bron
(e.g., \textit{work} together with \textit{bibliography}) or 
% work should be distinguished from influence/scientist
included labels which were seemingly unrelated (e.g., \textit{influence} and \textit{scientist}).
% werk is not with werken
They remarked on some titles which were not clustered together: \textit{Work} was not in the same cluster as \textit{works}, 
% separation on military and politic career
\textit{military career} and \textit{political career} belong to different clusters,
%why life and young years together but not death, beginjaren
\textit{life} and \textit{young years} were clustered together but \textit{death} and \textit{early years} were in two different clusters.
In brief, the clustering quality was acceptable to continue.

\textit{Aspect Classification.}
We evaluated the aspect classification in three ways: (1) Wikipedia classifier quality measured on test sets, (2) article classification statistics, and (3) rated classified snippets in a manual evaluation. 
For the aspect mining, we selected an absolute support of 100 examples per aspect to ensure enough examples for the subsequent classification, and a relative support of 0.05 to ensure relevance to the role. 
With that, we trained classifiers for 43 roles that had at least three different frequent aspects.
%1. Average Wikipedia Classification Results -> Table
We applied the classifiers to the Wikipedia test sets to measure the classification quality.
The results are reported in Tab.~\ref{tab_eval_classification} with additional statistics (avg. training data size, number of aspects). 
To look at the best and worst performing classifiers (ranked by macro precision to ensure reliable, precise classes), we evaluated five settings: Top-5 classifiers, Top-10 classifiers, Worst-5, Worst-10, and All classifiers. 
In brief, we concluded two thoughts: 
The more training data a classifier got, the better its performance was. 
Top-5 achieved a macro precision of 0.95, while Worst-5 still maintained a precision of 0.8, which we still consider acceptable.  
The recall was between 0.94 (Top-5) and 0.79 (Worst-10).
The number of trained samples was between 10k and 1.2k. 
However, the deviation was high, e.g., a deviation of 14k for 10k samples. 
A close look at histograms revealed some outliers, like the role \textit{person} with more than 150k samples. 
Overall, the classification quality was good.

\begin{table}[t]
\centering
\caption{Evaluation results of our article snippet classification. For each person, the number of used snippets, different roles,  snippets classified as belonging to one aspect, aspects, and classified snippets per aspect are reported.}
\begin{tabular}{lrrrclcc}
\label{tab_classified_snippets}
\textbf{Name} & \textbf{\#Sni.} & \textbf{\#Roles} & \textbf{\#Classified} & \textbf{\#Aspects} & \multicolumn{3}{c}{\textbf{Snippets/Aspects}} \\
\toprule
 &  &  &  & & Mean$\pm$STD & Min & Max \\
\midrule
W. Churchill & 48k & 15  & 47k & 92 & $508\pm1587$ & 1 & 12172 \\

Leopold III & 3192 & 6 & 1691 & 42 & $40\pm61$ & 1 & 332 \\    
Wilhelmina  & 40k & 1 & 231 & 5 & $46\pm24$ & 16 & 76 \\

Jannetje Schaft & 76 & 1 & 1 & 1 & $1\pm0$ & 1 & 1 \\
D. Eisenhower & 100k &  3 & 36k & 20 & $1780\pm6631$ & 1 & 30568 \\

Anne Frank  & 9 &  4 & 1 & 1 & $1\pm0$ & 1 & 1 \\

Frans Goedhart & 2995 & 2 & 1132 & 12 & $94\pm283$ & 1 & 1031 \\

Simon Vestdijk & 4989 & 3 & 3368 & 20 & $168\pm462$ & 1 & 2154 \\

F. Roosevelt & 80k & 7 & 40k & 50 & $799\pm3116$ & 1 & 21926 \\
\end{tabular}
\end{table}

%  2. Statistics of News article classification -> Table
Table~\ref{tab_classified_snippets} reports statistics on the actual classified news article snippets. 
For instance, Winston Churchill had up to 15 different roles yielding 47k classified snippets with 92 different role aspects in total. 
While some role aspects had up to 12k classified snippets, others had only one. 
Briefly, the number of classified snippets strongly differed between our test persons. 
Persons like Wilhelmina did not have a role concerning Wikipedia and were hence classified only as a \textit{person}.
%3. Hand-crafted evaluation of classified article snippets. -> Text
In our user interface, we show the best-classified snippets plus their summary.
That is why we ranked classified snippets by their classification probability and selected the top-5 per person, role, and aspect.
From this list with 557 snippets, we randomly sampled 100 entries (role, aspect, snippet) for a manual evaluation. 

Three persons rated each entry's correctness and gave explanations if they tagged an entry as incorrect.
Counting the majority votes, we obtained 62 correct and 38 incorrect entries with an inter-rater agreement of 0.33 (Krippendorff's $\alpha$~\cite{krippendorff1989content}) and 0.32 (Fleiss' $\kappa$~\cite{fleiss1971}). 
Discussing the reasons for the negative ratings revealed that, in many cases, the aspect applied was correctly classified, but the role did not fit.
For instance, some aspects like \textit{early life} were way too general to be specific for one role, and hence, deciding whether an \textit{early life} snippet belonged to the role \textit{politician}, \textit{member of the Parliament}, or \textit{writer} was impossible.
Annotators were uncertain about how to rate a statement about a person rather than an action performed by the person. 
Another encountered issue was distinguishing between pairs of roles which could belong together: \textit{journalist} -- \textit{writer}, \textit{minister} -- \textit{official}, \textit{writer} -- \textit{artist}, or \textit{historian} -- \textit{writer}.
Such a decision strongly influenced the rating of the aspect classification and oftentimes made raters disagree. 
Some snippets alone were not enough to rate an entry, e.g., if the award \textit{Karlspreis} is given to \textit{writers}. 

% facts/influence was unclear: statement of person or person doing something
%is journalist the same as writer $\rightarrow$ example 16
%is a minister an ambtenaar $\rightarrow$ example 20
%writer vs artist $\rightarrow$ example 27
%historian vs. writer

%snippets alone are sometimes not enough to decide $\rightarrow$ example 23, 32, 92 (Karlspreis)

%$\rightarrow$ too fine-grained in lower levels

\textit{Translation.}
We randomly sampled 100 snippets from all news articles. 
Two native Dutch speakers read the Dutch snippet and the corresponding translated English version. 
They rated the syntax of the translation (whether it reads well and is syntactically correct) and the factual correctness (whether the translated facts are still correct). 
For the syntax, the annotators' ratings for \textit{good-moderate-bad} were 54-28-47 and 38-47-15.
%rater one scored 54 as \textit{good} and 28 as \textit{moderate}, and 18 as \textit{bad}.
%Rater two scored 38 as \textit{good}, 47 as \textit{moderate}, and 15 as \textit{bad}.
The inter-rater agreement was 0.62 (Krippendorff's $\alpha$) and 0.39 (Fleiss' $\kappa$).
However, annotators often disagreed in rating a snippet as good or moderate. 
Counting good and moderate together as one class, we obtained an inter-rater agreement of 0.75 (Krippendorff's $\alpha$ and Fleiss' $\kappa$), indicating a fair agreement.
In brief, between 82-85\% of the translation snippets had a moderate or good syntax. 
Concerning factual correctness, for \textit{correct-incorrect} the ratings were 82-18 and 85-15.
%rater one scored 82 as correct and 18 as incorrect. 
%Rater two scored 85 as correct and 15 as incorrect. 
The inter-rater agreement was high, 0.85 (Krippendorff's $\alpha$ and Fleiss' $\kappa$).
Discussions with both raters revealed that in most cases, when a snippet was marked as factually incorrect, it was due to a minor error. 
The translation worked well with older Dutch, apart from some mistakes (such as \textit{'Duitschland’}, which was erroneously translated as \textit{‘Germanland’}). 
The translation also handled minor OCR errors or spaces.

\begin{table}[t]
\centering
\caption{Summarization evaluation results. The averaged readability scores, the averaged number of summary sentences, and the averaged reading time are shown per k.}
\begin{tabular}{crccccr}
\label{tab_eval_summaries}
\textbf{Summary@k}  & \textbf{\#Sent.} & \textbf{Flesch EN} & \textbf{Flesch NL}  & \textbf{Reading Time} & \textbf{Dale-Chall}\\
\toprule
5	&	7.7 $\pm$ 2.8	&	70.4 $\pm$ 8.8	&	57.3 $\pm$ 9.7	&	11.9 $\pm$ 3.8	&	9 $\pm$ 0.9	\\

10	&	10.4 $\pm$ 4.1	&	70.1 $\pm$ 10.1	&	57.1 $\pm$ 10.2	&	15.5 $\pm$ 4.8	&	8.8 $\pm$ 0.9	\\

20	&	13.7 $\pm$ 5.8 &	69.8 $\pm$ 9.0	&	56.4 $\pm$ 9.0	&	19.4 $\pm$ 5.6	&	8.8 $\pm$ 0.7	\\

30	&	15.0 $\pm$ 5.9	&	70.2 $\pm$ 8.4	&	56.6 $\pm$ 9.4	&	21.0 $\pm$ 6.4	&	8.7 $\pm$ 0.8	\\

40	&	15.6 $\pm$ 6.4	&	70.5 $\pm$ 8.0	&	57.3 $\pm$ 8.8	&	21.7 $\pm$ 6.8	&	8.7 $\pm$ 0.7	\\

50	&	16.1 $\pm$ 6.4	&	71.3 $\pm$ 8.0	&	57.9 $\pm$ 8.3	&	22.0 $\pm$ 6.9	&	8.7 $\pm$ 0.7	\\

\end{tabular}
\end{table}
 
\textit{Summarization.}
We evaluated the summarization through (1) automated readability scores and (2) a manual evaluation.
We only summarized aspects of roles that had at least five classified snippets per person, otherwise, we did not have enough information to show.
In addition, we summarized the most probable 20 article snippets based on classification probability as the multi-document summarization model could only process 4096 tokens as its input and will truncate otherwise. 
With that, we generated 208 summaries in total. 
%2. Automated readability scores -> Table
Table~\ref{tab_eval_summaries} reports the following averaged measures:
The Flesch readability index~\cite{Flesch1948ANR} quantifies reading ease based on word and sentence length and is language-specific. 
Scores between 50 and 60 indicate fairly difficult text, while scores between 70 and 80 indicate fairly easy text.
The reading time indicates the seconds required to read a text, each character taking 14.69ms~\cite{reading_time}.
The new Dale-Chall~\cite{chall1995readability} score gives the reading level of a text as a grade indicating the familiarity of persons from that grade with a list of words. 
Scores from 8.0 to 8.9 correspond to an 11th/12th-grade student's reading level.
We also tested a different number of selected snippets to summarize \textit{k}, however, except for the number of generated sentences and the required reading time, the scores did not deviate much. 
%3. Hand-crafted summary evaluation -> Text
We randomly sampled ten summaries plus the 20 snippets used to generate them for three raters. 
%Each rater commented on the readability (bad, average, good) and factual correctness concerning the original snippets. 
Readability on a \textit{good-average-bad} scale was rated as 0-10-0, 0-10-0 and 1-4-5.
%Rater one and two rated all summaries as average, whereas rater three rated one as good, four as average, and five as bad.
The generation quality was hence acceptable. 
From their discussion of the results we found the following: First, if some snippets supported parts of the summary, they were nearly cited verbatim. 
Some snippets of different articles were (nearly) identical.
Temporal information in the summaries on dates was often bad because the dates were wrong or messed up. 
Some summaries included hard context breaks between sentences. 
Moreover, we observed major issues with factual correctness due to hallucinations. 
Phrases like \textit{``The New York times reports, click here for more''} were generated but not included in the articles.
Further, the model also introduced additional, and often wrong, facts about persons, e.g., dates, events, and actions.
We assume that such facts and phrases were already learned in PRIMERA's pre-training and fine-tuning for news.
However, hallucinated facts in summaries were a major issue. 
%A similar observation was made by Zhang et al.~\cite{zhang-etal-2023-enhancing} who reported a factual consistency between 51-55\% in a comparable setting (trained on news, tested on another). 
A comparable setting (trained on news, tested on other data) found 51-55\% factual consistency~\cite{zhang-etal-2023-enhancing}.

%read quite well, moderate quality, partially factually correct.
%lots of hallucination: The New York Times reports, click here for more/pictures

%temporal things are always bad $\rightarrow$ example 4

%connections between unrelated things are spun/unrelated things $\rightarrow$ example 9

%content hard broken

%not all snippets contributed to the summary

%if fact from article, it is nearly verbatim cited

%if freely generated, the text sounds quite good

%\todo{Hermann: writes algorithm to evaluate the verbatim parts via sentence alignment}

%\textit{Runtimes.}
%We performed every operation on our server having ...
%Every operation was performed on a single CPU/GPU.
%The pre-processing of Wikipedia took a few ours. 
%Fine-tuning the text classification models took roughly two days.
%\todo{measure}Classifying articles of one person -> compute per no. of articles. Test, 3x repetition
%Summarizing articles without translation cache -> measure time

\textit{User Interviews.}
We conducted five independent 30-minute interviews with employees of the KB.
The interview partners consented to take part in the study voluntarily and have their voices recorded. 
They were made aware that they could stop and drop out of the interview at any time without consequences. 
The process (a mail to the investigator) for later deleting user-specific data was also explained. 
A week before our semi-structured voice-recorded one-on-one interviews in Dutch took place, participants received an email with a video (\url{https://www.youtube.com/watch?v=0GzIydjts2E}) 
explaining our prototypical system and its URL. Each interview tackled the same guide questions concerning general thoughts, encountered problems, (un)clear elements, helpfulness of the system and components (aspects and summaries), and suggested changes.

%were posed. 
%In the following, we discuss the answers to these questions:
%\begin{inparaenum}
 %   \item What are your general thoughts regarding the system?
  %  \item Where did you encounter problems? What was unclear?
   % \item What did you like/immediately understand?
    %\item How do you think this prototype could help users of the KB?
    %\item What do you think about the roles of a person?
    %\item What do you think about the summaries for the articles?
    %\item Which changes would you suggest in order to make the system more suitable/helpful to users?
    %\item Anything else you want to add or ask?
%\end{inparaenum}

\textit{Results and Findings.}
The full questions plus answers are available in our GitHub repository. 
In general, the interviewees were enthusiastic about the interface. They found it well-arranged and clear. The website immediately provided a lot of information and context about the person in question. Some other remarks were that the interface worked intuitively and that clustering articles per subject was a plus point. The option to be directly referred to the complete articles on Delpher was also mentioned positively. The interviewees believed that a website like this could definitely help certain users (such as researchers), mainly because they immediately get some context about a person instead of only a list of articles. However, they all agreed that some human input was still needed to refine the system's output. The interviewees also provided feedback on the various aspects of the system. They found the roles interesting and a good way to immediately provide information about the person. However, they all had some difficulty in understanding how the roles were chosen as they noticed a lot of overlap between different roles. This led to the question of why these have not been merged (such as the roles \textit{politician} and \textit{politician by party}). Opinions were divided on the aspects. Some found the distinction useful, while others wondered why not all articles belonging to one role were grouped together. They agreed that the multiple labels (clustered section titles of Wikipedia) shown above every aspect should be condensed for clarity for two reasons: The number of labels is unbalanced between aspects, and  the sections with many labels caused some confusion, e.g., the aspect with the labels \textit{background}, \textit{biography}, etc. appeared under every role. The interviewees expected it to only belong to the role \textit{person}. Summaries really posed major issues and worried the interviewees. All unanimously agreed that summaries containing incorrect facts are highly problematic for a library. 
Some also wondered whether the summary added value.

\section{Conclusion}
In this work, we demonstrated how a digital library can implement an aspect-driven access path to its news collection. 
We used Wikipedia to bypass the curation of domain-specific and cost-intensive training data. 
Moreover, our evaluation verified the method's effectiveness on real-world data and the system's value in practice.  
However, there is still room for improvements, e.g., finding suitable labels for a section cluster, showing and summarizing diverse snippets, and highlighting connections between people.
For instance, we could better cater to the requirements of the KB by battling hallucinations in summaries by either fact-checking each sentence against the input summaries and removing unsupported ones or by using an extractive summarization approach~\cite{DBLP:conf/emnlp/LiuL19,extractiveSummarization}.

%\subsubsection{Acknowledgements} Supported by Deutsche Forschungsgemeinschaft (DFG, German Research Foundation): PubPharm – Specialized Information Service for Pharmacy (Gepris 267140244).

\bibliographystyle{splncs04}
\bibliography{bibliography}
\end{document}